\documentclass[reprint,amsmath,amssymb,aps]{revtex4-2}
\usepackage{graphicx}
\usepackage{dcolumn}
\usepackage{bm}
\usepackage{wrapfig}
\usepackage{soul,xcolor}
\usepackage{physics} 
\usepackage{mathtools}
\usepackage{array}
\usepackage{comment}
\usepackage{makecell}
\begin{document}

\title{Tetrahedral rotations in the alkaline-earth metal orthovanadates}

\author{Amartyajyoti Saha,$^{1,2}$ Turan Birol$^1$}
\affiliation{$^1$Department of Chemical Engineering and Materials Science, University of Minnesota}
\affiliation{$^2$School of Physics and Astronomy, University of Minnesota}


\begin{abstract}
The alkaline-earth metal orthovanadate Sr$_3$(VO$_4$)$_2$ with the palmierite structure is reported to host a dielectric anomaly as well as a structural phase transition above the room temperature. With V$^{5+}$ ions and tetrahedral oxygen coordination, the crystal structure of this compound is not studied in detail from first principles yet. In this work, we perform a detailed analysis of the crystal structure and instabilities of $M_3$(VO$_4$)$_2$ ($M = \text{Ca, Sr, Ba}$) orthovanadates with the palmierite structure using first principles density functional theory. We find that as the $M^{2+}$ cation size decreases, a significant structural distortion that changes the symmetry from $R\bar{3}m$ to $C2/c$ emerges. This change is accompanied with a rotation of the oxygen tetrahedra. Our calculations also indicate that the polar instability in these compounds is suppressed by these tetrahedral rotations.   
\end{abstract}

\maketitle

\section{\label{sec:intro}Introduction}

Perovskites with the chemical formula $AB$O$_3$ are one of the most studied families of oxides, in no small part because of their flexibility to host a large number of different cations in both $A$ and $B$ sites. In many early transition metal perovskite oxides, for example the titanates, the details of the electronic structure near the Fermi level is determined by the $B$ cation, while the $A$-site cation modifies the crystal structure mostly through steric effects. For example, while all three of them are band insulators, CaTiO$_3$ is a dielectric with large octahedral rotations, SrTiO$_3$ is a quantum paraelectric, and BaTiO$_3$ is a room-temperature ferroelectric \cite{Rabe2007}. Similarly, the rare-earth titanates $R^{3+}$TiO$_3$ are all magnetic Mott insulators, but changing the rare earth cation on the $A$-site modifies the parameters of the crystal structure just enough to modify the magnetic exchange interactions, and lead to a ferromagnetic-antiferromagnetic transition \cite{Greedan1985, Mochizuki2004}. 

While the perovskite structure allows different structural distortions, such as octahedral rotations \cite{Glazer1972}, to accommodate cations with a wide range of radii, there is a finite range of $B$-site cation sizes that can fit in the $B$O$_6$ octahedral units. In particular, for smaller transition metal cations, the octahedral anion-coordination becomes unfavorable due to Pauling's first rule \cite{Pauling1929}. As a result, for transition metals that can have different valence states, such as vanadium, the highest valence state is often not stable in the perovskite configuration. While, for example CaV$^{4+}$O$_3$, SrV$^{4+}$O$_3$, and BiV$^{3+}$O$_3$ all exist in perovskite form, to the best of our knowledge, no $A^{1+}$V$^{5+}$O$_3$ perovskite exists. CsVO$_3$ and KVO$_3$ instead form structures where each V ion is surrounded by four oxygen anions, forming a tetrahedron, consistent with the Pauling's rules \cite{Luo2021, Nakajima2015}. 

There is an increasing interest in vanadium oxides due to, for example, the promising optical properties of BiVO$_3$ \cite{Praveen2017}, the transparent conducting properties of SrVO$_3$ and CaVO$_3$ \cite{Zhang2016}, photocatalytic properties BiVO$_4$, as well as the recent prediction of antiferroelectricity in KVO$_3$ \cite{Aramberri2020}. However, the first principles work on the V$^{5+}$ containing compounds is rarer than the work on systems with V$^{4+}$. It is therefore important to perform theoretical characterization of the structure and properties of different vanadate compounds hosting V$^{5+}$.

In this work, we present a first principles analysis of the crystal structure and possible transitions of the alkaline-earth metal orthovanadates $M_3$(VO$_4$)$_2$ ($M = \text{Ca, Sr, Ba}$) with the palmierite-related structures.
To the best of our knowledge, there is no systematic study of the crystal and electronic structures of these orthovanadates, even though they have been known for a long time. For example, Ba$_3$(VO$_4$)$_2$ was already well characterized in 1970 \cite{Susse1970}, and Ca$_3$(VO$_4$)$_2$ in a different but related structure is known to be a room temperature pyroelectric since 1973 \cite{Gopal1973, Glass1977}. 
The dielectric properties of palmierite Ba$_3$(VO$_4$)$_2$ have also been studied in the microwave range more recently \cite{Umemura2006}, and Sr$_3$(VO$_4$)$_2$ with the same crystal structure is reported to have a dielectric anomaly signaling a ferroelectric phase transition by multiple groups \cite{Pati2013, Pati2015, Batista2017}. 
In addition to their non-toxic nature and promising ferroelectric properties, these materials also exhibit interesting optical and transport properties \cite{Pati2013,Parhi2008,Lim2013}. In particular, strontium and barium orthovanadates, Sr$_3$(VO$_4$)$_2$ and Ba$_3$(VO$_4$)$_2$, are known to exhibit intense rare earth activated luminescence and can be used as luminophores and host materials for lasers. 

First-principles calculations can identify the low temperature crystal structure, as well as metastable high-temperature phases in crystalline materials. In this work, we use a combination of density functional theory (DFT) including lattice dynamics calculations, evolutionary structure prediction algorithms, and group theory to identify the possible structural phases of $M_3$(VO$_4$)$_2$ ($M=\text{Ca, Sr, Ba}$) with the palmierite structure. We show that the lowest energy structures with a single formula unit are derived from the palmierite structure for all three of these compounds, and that all of them have dynamical lattice instabilities with varying strengths. The strongest of these lattice instabilities is in the form of rotation (or tilt) of the oxygen tetrahedra, in a way that resembles the oxygen octahedral rotations commonly observed in perovskite oxides \cite{Glazer1972, Lufaso2004}. The lowest-energy crystal structure of Sr$_3$(VO$_4$)$_2$ is significantly lower in energy than the undistorted palmierite structure, which suggests that a detailed structural characterization of this compound should reveal tetrahedral rotations. While we identify a polar instability in this compound, the polarization is suppressed by these rotations, and hence is absent from the ground state, which suggests an extrinsic origin to the dielectric anomalies reported in Sr$_3$(VO$_4$)$_2$. 

This paper is organized as follows: In Sec.~\ref{sec:methods}, we introduce the details of our computational methods. In Sec.~\ref{sec:results}, we discuss the earlier observations of the crystal structures of these compounds, followed by the results of our evolutionary structure search, and lattice dynamics calculations. The DFT calculations are supported by a Landau free-energy analysis in Sec.~\ref{sub:landau} and a bond valence analysis in Sec.~\ref{sub:bondvalence}. We discuss the details of the electronic structure in Sec.~\ref{sub:electronic}. We conclude with a summary in Sec.~\ref{sec:conclusion}.


\section{\label{sec:methods}Methods}
First principles DFT calculations were performed using the Vienna \textit{ab initio} simulation package (\textsc{vasp}), which uses projector augmented wave formalism \cite{Kresse199607,Kresse199610,Blochl1994,Kresse1999}. 
The revised Perdew-Burke-Ernzerhof generalized gradient approximation for solids (PBEsol) was used to approximate the exchange correlation energy \cite{Perdew2009} unless stated otherwise. 
To confirm that there is no other crystal structure with the same stoichiometry that is thermodynamically more favorable, i.e., the palmierite structure is the valid high-symmetry reference structure for the orthovanadates, structural prediction was performed using an \textit{ab initio} evolutionary search algorithm as implemented in the \textsc{uspex} package \cite{Glass2006,Oganov2006} in conjunction with \textsc{vasp}. In these calculations, a large number of initial structures of different symmetries were randomly generated and relaxed using DFT.
Subsequent generations were generated by introducing ``mutations'' to earlier ones, as well as by adding new random structures.
A total of about 600 structures were considered for barium and strontium orthovanadates individually, while about 800 structures for calcium orthovanadate were considered.

For each compound, few of the lowest energy structures obtained from the evolutionary search were further relaxed on a shifted Monkhorst-Pack \cite{Monkhorst1976} uniform $k$-point sampling with a resolution of 0.1 {\AA}$^{-1}$ or better. 
In order to look for instabilities, phonon calculations were performed using the \textit{direct method} as implemented in the \textsc{phonopy} package \cite{Togo2015} on a $2\times2\times2$ supercell.
To accurately predict the band gap, band structure calculation was performed using improved Heyd-Scuseria-Ernzerhof hybrid function for solid (HSEsol) \cite{Schimka2011,Heyd2003} on a $4\times4\times4$ $k$-point grid, which was sufficient to obtain converged band structures. \footnote{See Supplemental Material for further details on the $k$-point convergence of band structure, the details of the anion repulsion energy calculation, and the higher energy structure predicted for Ca$_3$(VO$_4$)$_2$ by the evolutionary structure prediction calculation.}
All calculations were performed with an energy cutoff of 500 eV for the plane waves. 
A force per atom of $5\times10^{-4}$ eV/{\AA} and an energy of $10^{-7}$ eV were used as the convergence criteria. 
Spin-orbit coupling was not taken into account in any of the calculations reported in this paper, but it was confirmed to cause no qualitative or significant quantitative difference.

The \textsc{wannier90} package \cite{Mostofi2008,Marzari2012} was used to calculate maximally-localized Wannier functions (MLWFs). The distorted structures with specific irreducible representations (irreps) of each unstable mode were determined with \textsc{isodistort} \cite{ISODISTORT,Campbell2006}.
The Landau free-energy expressions were obtained using \textsc{invariants} \cite{INVARIANTS,Hatch2003}, which generates all the invariant polynomials of possible order parameter components in a given space group. The Bilbao Crystallographic Server was used as a reference for group theory tables \cite{Aroyo2006}. Crystal structure visualization was carried out using the \textsc{vesta} software \cite{Momma2011}.


\section{\label{sec:results} Results and Discussion}

\subsection{\label{sub:expt} Experimentally observed structures}

\begin{figure}
    \includegraphics[width=0.98\linewidth]{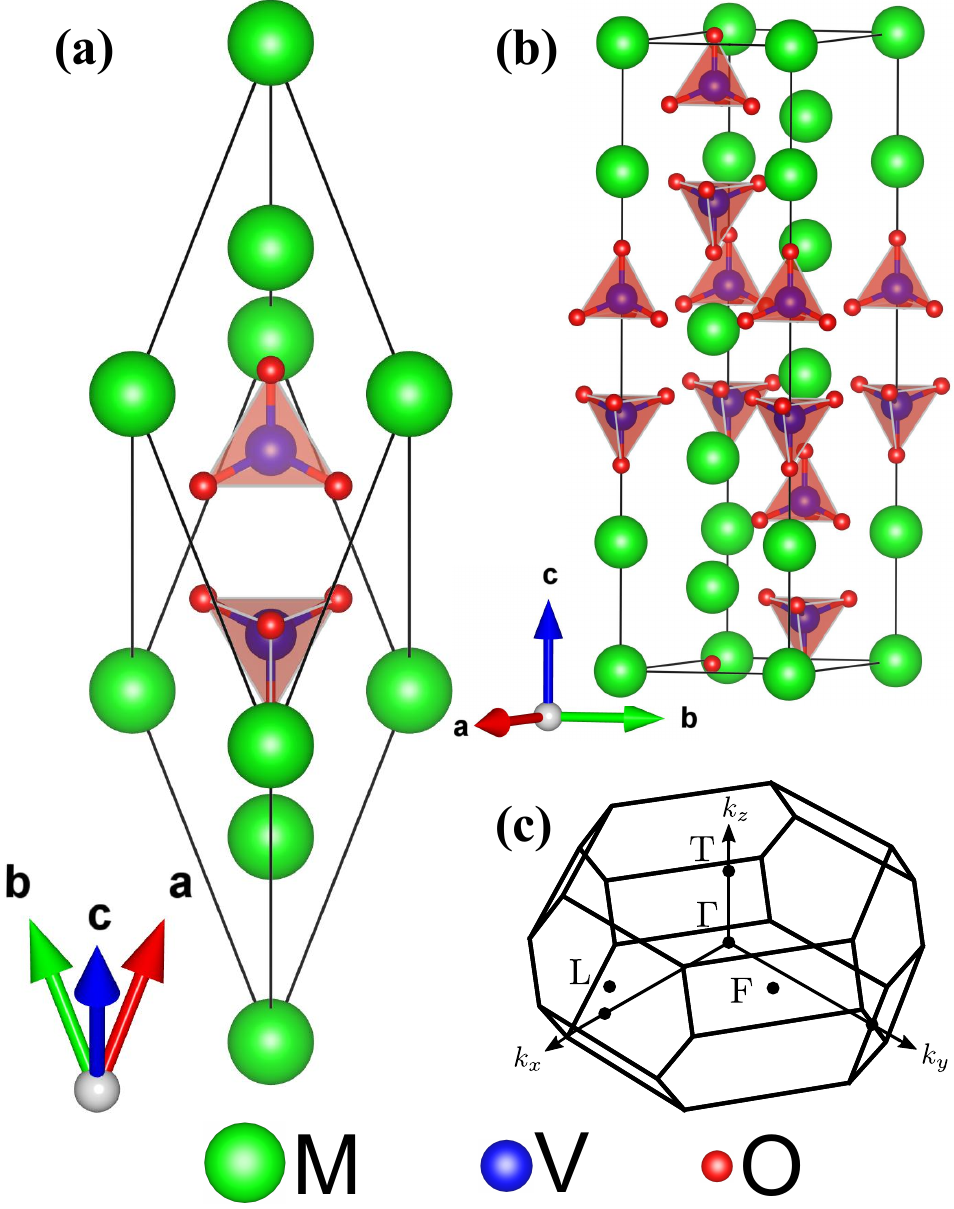}
    \caption{Crystal structure of the orthovanadates in the high-symmetry $R\bar{3}m$ structure. (a) Primitive and (b) conventional unit cells are shown with the (c) Brillouin zone of the primitive cell. The [111] direction of the primitive cell corresponds to the [001] direction of the conventional cell.}
    \label{fig:structure}
\end{figure}

There are several experiments that performed structural characterization on the alkaline-earth metal orthovanadates. (See, for example, Ref.'s~\cite{Susse1970,Gopal1973,Grzechnik2002,Grzechnik1997}). Ba$_3$(VO$_4$)$_2$ and Sr$_3$(VO$_4$)$_2$ are observed to have the palmierite structure at room temperature with $R\bar{3}m$ symmetry \cite{Grzechnik1997, Susse1970}, as shown in Fig.~\ref{fig:structure}.
This structure is derived from the the structure of K$_2$Pb(SO$_4$)$_2$ palmierite \cite{Tissot2001}. 
It has [$M_{(1)}$(VO$_4$)$_2$]$^{4-}$ layers linked into the crystalline structure by $M_{(2)}^{2+}$ cations.
The isolated VO$_4$ groups are centered on V sites with $3m$ symmetry, and hence are slightly distorted from the ideal tetrahedral geometry. 
Among the four V--O distances, one is shorter than the other three.
There are two different alkaline-earth metal sites, with different symmetries ($3m$ and $\bar{3}m$) and different coordination environments. 

Ca$_3$(VO$_4$)$_2$ is observed to have a different structure with $R3c$ \cite{Gopal1973,Glass1977} symmetry in ambient conditions and $C2/m$ symmetry under pressure \cite{Grzechnik2002}.
The $R3c$ structure has a primitive unit cell consisting seven formula units, and partial occupations. In this study, we focus on a yet-to-be-synthesized palmierite polymorph of Ca$_3$(VO$_4$)$_2$ to predict the structural trends as a function of changing alkaline-earth ion radius in the Ba--Sr--Ca series of orthovanadates. 

An open question about the structure of Sr$_3$(VO$_4$)$_2$ is whether it is polar. 
The dielectric anomaly is reported to take place well above the room-temperature \cite{Pati2015}, which would require the room temperature space group to be noncentrosymmetric. 
This is inconsistent with $R\bar{3}m$ space group, which has inversion symmetry. First principles calculations can shed light on this seeming inconsistency by predicting the unstable lattice modes (or lack thereof) in the centrosymmetric structure with $R\bar{3}m$ symmetry, thereby providing novel insights into the polarity of Sr$_3$(VO$_4$)$_2$. 



\subsection{\label{sub:uspex} Evolutionary structure prediction}

To find a thermodynamically stable crystal structure without introducing any bias to the prediction, we started by performing an evolutionary structure search using \textsc{uspex}. We considered a unit cell consisting of one formula unit and a population of randomly generated 30 initial structures considering all possible symmetries. On each generation, our calculations kept the best 5 structures from the previous generations and generated 45 new structures. To generate these new structures, the fittest 70\% structures of the earlier generations are modified in various ways, while in every generation, a few randomly generated structures are also considered.
We then relaxed the structures in several steps with increasing numerical accuracy and decreasing constraints.
Structural prediction was considered to be completed if the ground-state structure did not change for 10 consecutive generations.

This evolutionary search predicts a Ba$_3$(VO$_4$)$_2$ structure with $R\bar{3}m$ symmetry and Ca$_3$(VO$_4$)$_2$ and Sr$_3$(VO$_4$)$_2$ structures with slightly lower $R\bar{3}$ symmetry (Fig.~\ref{fig:structure})
Within the first three generations for each compound. For Ca$_3$(VO$_4$)$_2$, the evolutionary search also identifies a different metastable structure with $C2/m$ symmetry, which we include in the Supplemental Material \cite{Note1}. This structure is 76 meV/f.u.  (formula unit) higher in energy than the lowest-energy structure, and we do not explore it any further. The evolutionary searches for Ba$_3$(VO$_4$)$_2$ and Sr$_3$(VO$_4$)$_2$ do not yield any other likely metastable structure not related to the $R\bar{3}m$ one via a dynamic instability.
The $R\bar{3}m$ structure is the palmierite structure previously reported, and consists of disconnected VO$_4$ tetrahedra. 
The $R\bar{3}$ structures can be obtained from the $R\bar{3}m$ structure with a small rotation of the VO$_4$ tetrahedra.
We manually obtain the $R\bar{3}m$ structure for Ca$_3$(VO$_4$)$_2$ and Sr$_3$(VO$_4$)$_2$ (respectively 176 and 24 meV/f.u. higher in energy from $R\bar{3}$ structure) in order to investigate the emergence of structural instabilities in all three compounds together.

\subsection{\label{sub:phonon} Lattice dynamics of $\mathbf{\textit{M}_3}$(VO$\mathbf{_4}$)$\mathbf{_2}$}

\begin{figure}[ht]
    \centering
    \includegraphics[width=0.98\linewidth]{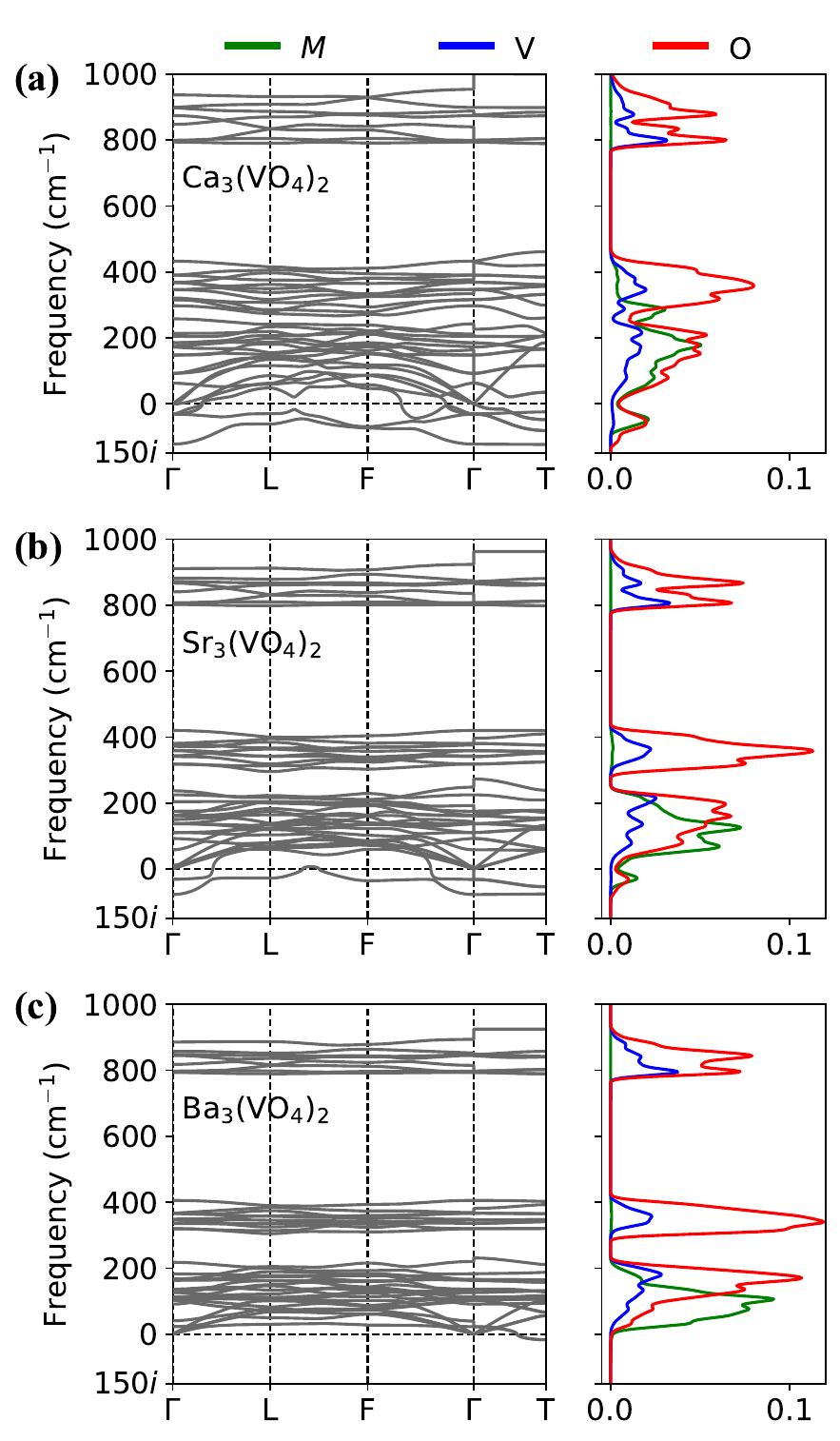}
    \caption{Phonon dispersion curves and the corresponding phonon partial DOS for the orthovanadate $R\bar{3}m$ structure. The atomic characters of the phononic wavefunctions were obtained from atomic spheres, renormalized to account for interstitial densities, and denoted with  green ($M$), blue (V), and red (O) colors. DOS suggests that $R\bar{3}m$ structure of calcium and strontium orthovanadate have oxygen and $M$ site instabilities. The units of phonon DOS is cm/f.u..}
    \label{fig:phonon}
\end{figure}

In order to identify possible structural transitions and map out possible metastable phases not captured by the evolutionary structure prediction, we performed first principles lattice dynamics calculations on the parent $R\bar{3}m$ structure of each compound. The phonon dispersion and the atom projected phonon density-of-states (PDOS) are shown in Fig.~\ref{fig:phonon}.
Surprisingly, all three compounds, including Ba$_3$(VO$_4$)$_2$ display lattice instabilities. Lattice instabilities at high-symmetry points, which are indicated by imaginary frequencies, are listed in Table \ref{tab:phonon} along with their corresponding irreps.

\begin{table}[]
    \centering
    \begin{tabular}{ c c c c }
    \hline
    \hline
    $k$-points  
    & \begin{tabular}{c} Ca$_3$(VO$_4$)$_2$ \\ ($\times i$ cm$^{-1})$ \end{tabular} 
    & \begin{tabular}{c} Sr$_3$(VO$_4$)$_2$ \\ ($\times i$ cm$^{-1})$ \end{tabular} 
    & \begin{tabular}{c} Ba$_3$(VO$_4$)$_2$ \\ ($\times i$ cm$^{-1})$ \end{tabular} \\
    \hline
    \hline
    
    $\Gamma$  
    &  \begin{tabular}{c}
      120 ($\Gamma_2^+$) \\  30 ($\Gamma_3^-$) \\ 18 ($\Gamma_1^-$)
    \end{tabular}  
    &  \begin{tabular}{c}
      76 ($\Gamma_2^+$)  \\  30 ($\Gamma_3^-$)
    \end{tabular} 
    & -- \\
    \hline
    
    $L$  
    &  \begin{tabular}{c}
      60 ($L_1^-$)  \\  30 ($L_1^-$)
    \end{tabular} 
    &  27 ($L_1^-$) 
    & -- \\
    \hline
    
    $F$ 
    &  \begin{tabular}{c}
      72 ($F_1^-$)  \\  68 ($F_2^+$)
    \end{tabular}  
    &  34 ($F_1^-$) 
    & -- \\
    \hline
    
    $T$ 
    &  \begin{tabular}{c}
      120 ($T_1^-$) \\  81 ($T_3^-$) \\ 48 ($T_3^+$) \\ 11 ($T_2^+$)
    \end{tabular} 
    &  \begin{tabular}{c}
      76 ($T_1^-$)  \\  53 ($T_3^-$)
    \end{tabular}  
    & 15 ($T_3^-$) \\
    
    \hline
    \end{tabular}
    \caption{Magnitude of instabilities of the $M_3$(VO$_4$)$_2$ R$\bar{3}$m structure at high-symmetry points. The corresponding irreps are shown inside brackets.}
    \label{tab:phonon}
\end{table}

\begin{table}[]
    \centering
    \begin{tabular}{c}
    Ca$_3$(VO$_4$)$_2$ \\
    \end{tabular}\\
    \begin{tabular}{c c c c c c}
    \hline
    \hline
    Atom &  Wyck. Pos. & Multiplicity & $\Gamma_2^+$  &  $T_1^-$  &  $T_3^-$ \\
    \hline
    Ca & c & 6 & 0 & 0 & 0.237\\
    Ca & a & 3 & 0 & 0 & 0.296\\
    V & c & 6 & 0 & 0 & 0.229\\
    O & c & 6 & 0 & 0 & 0.481\\
    O & h & 18 & 0.408 & 0.408 & 0.195\\
    \hline
    \end{tabular}\\
    \begin{tabular}{c}
    \\Sr$_3$(VO$_4$)$_2$ \\
    \end{tabular}\\
    \begin{tabular}{c c c c c c}
    \hline
    \hline
    Atom &  Wyck. Pos. & Multiplicity & $\Gamma_2^+$  &  $T_1^-$  &  $T_3^-$ \\
    \hline
    Sr & c & 6 & 0 & 0 & 0.098\\
    Sr & a & 3 & 0 & 0 & 0.620\\
    V & c & 6 & 0 & 0 & 0.127\\
    O & c & 6 & 0 & 0 & 0.472\\
    O & h & 18 & 0.408 & 0.408 & 0.141\\
    \hline
    \end{tabular}
    \caption{Magnitude of eigenvector components for the three largest unstable modes of $R\bar{3}m$ Ca$_3$(VO$_4$)$_2$ and Sr$_3$(VO$_4$)$_2$. For comparison between $T$ and $\Gamma$ irreps, all eigenvectors are normalized in the primitive unit cell. The directions of the displacements are shown in Fig.~\ref{fig:rotation}.}
    \label{tab:eigenvector}
\end{table}

Ba$_3$(VO$_4$)$_2$ in the $R\bar{3}m$  structure has only a weak instability at the $T$ point, which transforms as the $T_3^-$ irrep. This is a two-dimensional (2D) irrep, and the corresponding distortion leads to structures with either $C2/c$ or $C2/m$ symmetry depending on what its direction is.
However, the energy gain in those structures is as small as $\sim$0.5 meV/f.u.. While a direct conversion between the DFT energy gain due to a structural change and the transition temperature at which the structural distortion would emerge is not possible, such a small change is expected to give rise to a transition at a few Kelvin at most. So, we conclude that Ba$_3$(VO$_4$)$_2$ has the $R\bar{3}m$ symmetry at room temperature, and is likely to remain so down to few Kelvin. 

\begin{figure}[ht]
    \centering
    \includegraphics[width=0.98\linewidth]{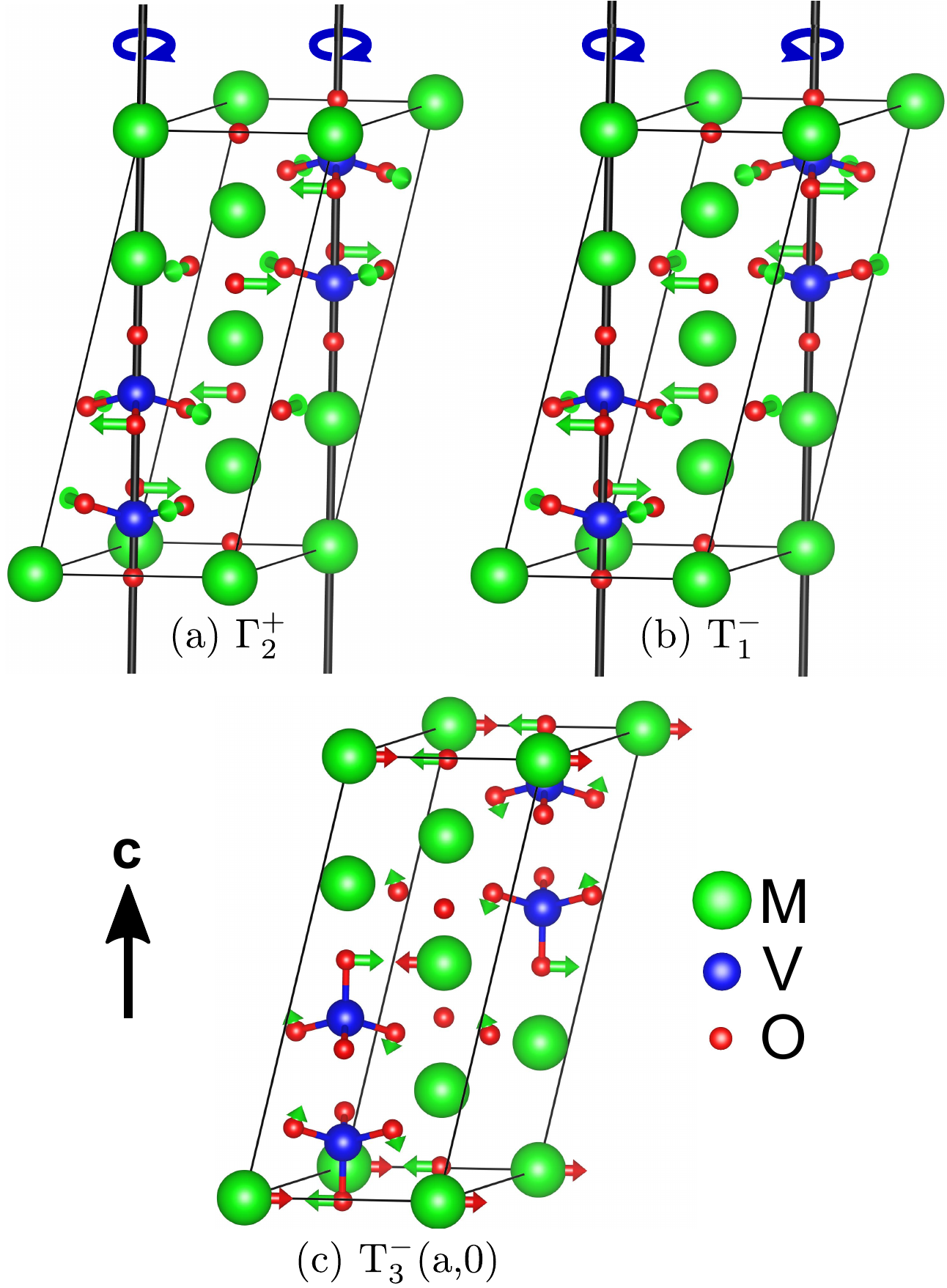}
    \caption{The sketch of the three different distortions considered  in the high-symmetry $R\bar{3}m$ structure of $M_3$(VO$_4$)$_2$. The distortions are shown in a supercell with two formula units of Ca$_3$(VO$_4$)$_2$ which is commensurate with the $T$ $k$-vector. (a) The $\Gamma_2^+$ irrep involves a tetrahedral rotation of the VO$_4$ unit about the $c$ axis of a conventional cell. (b) The $T_1^-$ irrep also represents the same tetrahedral rotation involving oxygens at the $h$ Wyckoff position, except the rotations are in the opposite direction in the neighboring unit cells. (c) The $T_3^-$ irrep causes the oxygen at the $c$ cite and the M atom at the $a$ cite to move closer to each other. This displacement also causes a tilt in the VO$_4$ tetrahedra.}
    \label{fig:rotation}
\end{figure}

Ca$_3$(VO$_4$)$_2$ and Sr$_3$(VO$_4$)$_2$ have much stronger instabilities than Ba$_3$(VO$_4$)$_2$, especially at $\Gamma$ and $T$ points, as shown in Figs.~\ref{fig:phonon}(a) and \ref{fig:phonon}(b).
The three strongest instabilities for both the compounds are caused by the same $\Gamma_2^+$, $T_1^-$, and $T_3^-$ irreps, and henceforth we focus on these instabilities' effect on the $R\bar{3}m$ structure.

The displacements due to these three unstable modes are shown in Fig.~\ref{fig:rotation}, and their displacement patterns (the components of the dynamical matrix eigenvectors) are listed in Table \ref{tab:eigenvector}. 
The one-dimensional (1D) irreps $\Gamma_2^+$ and $T_1^-$ correspond to the largest instabilities in the structure, and they cause rotations of the VO$_4$ units around the $c$ axis of the conventional unit cell as shown in Figs.~\ref{fig:rotation}(a) and \ref{fig:rotation}(b). The rotations are in the same direction for the $\Gamma_2^+$ irrep. For the $T_1^-$ irrep, the nearest-neighbor VO$_4$ units along the $ab$ plane have opposite  directions of rotation around the $c$ axis. 
The distortion corresponding to the $T_3^-$ irrep causes the oxygen at Wyckoff position $c$ and the $M$ atom at Wyckoff position $a$ to move towards each other. This displacement also causes a tilt in the VO$_4$ tetrahedra, as shown in Fig.~\ref{fig:rotation}(c).

\begin{figure}[ht]
    \centering
    \includegraphics[width=0.98\linewidth]{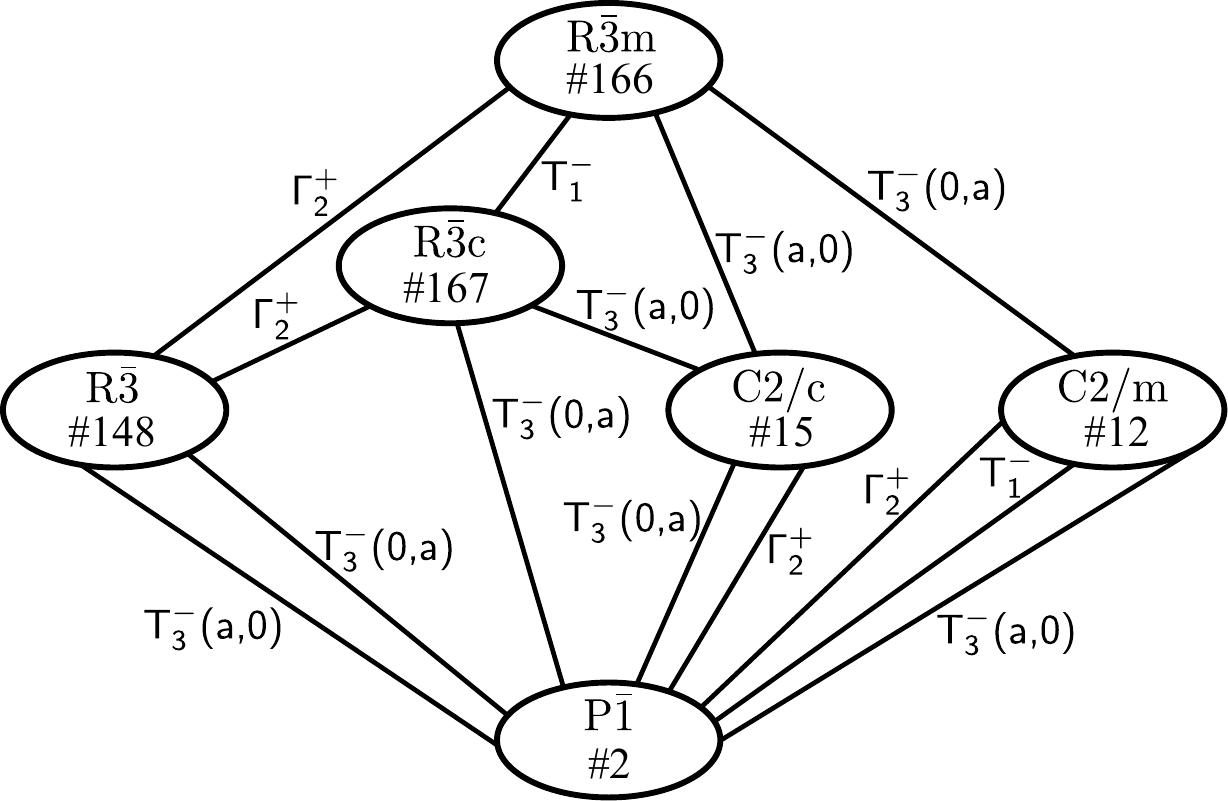}
    \caption{Group table showing the relevant structural phase transitions between the hexagonal $R\bar{3}m$ and the monoclinic $C2/c$ structures. The space group numbers are also shown below the Hermann-Mauguin notation. Other possible structural phase transitions are also shown with the corresponding symmetry of the lattice distortion.}
    \label{fig:group}
\end{figure}

To obtain the ground-state structure of Sr$_3$(VO$_4$)$_2$ and Ca$_3$(VO$_4$)$_2$, we looked into all possible combinations of these instabilities, which are listed in Fig.~\ref{fig:group}. Note that none of the groups listed in this figure, including those which have a combination of different structural distortions, are noncentrosymmetric. Hence, our first principles results suggest that stoichiometric Sr$_3$(VO$_4$)$_2$ is not ferroelectric. 

Of the distorted structures, the highest symmetry state where $\Gamma_2^+$, $T_1^-$ and $T_3^-$ can coexist is a two formula-unit structure with $P\bar{1}$ symmetry. 
In order to incorporate all the instabilities, we relaxed the $P\bar{1}$ structure, and found that the instability from the $\Gamma_2^+$ irrep vanishes and the resulting structure has $C2/c$ symmetry.
This structure incorporates instabilities from the $T_1^-$ and $T_3^-$ irreps and leads to an energy gain  of 47 meV/f.u. and 263 meV/f.u. for Sr$_3$(VO$_4$)$_2$ and Ca$_3$(VO$_4$)$_2$, respectively. These energy gains are larger than the energy scale of room temperature, which means that it is likely that these materials would have the $C2/c$ symmetry structure even at room temperature.

It is important to note that the evolutionary structure search we performed in Subsec.~\ref{sub:uspex} did not predict the structure with $C2/c$ symmetry because it contains two formula units of $M_3$(VO$_4$)$_2$.
The lowest energy structure that doesn't enlarge the unit cell, the $R\bar{3}$ structure which arises from the $\Gamma_2^{+}$ irrep, is correctly predicted by the evolutionary search. While there is a polar $\Gamma_3^-$ instability in both Sr and Ca compounds, it gives rise to a much smaller energy gain (merely 2  and 36~meV/f.u., respectively) than the $\Gamma_2^+$ or the $T$-point structural distortions that lead to the $C2/c$ structure.
Additionally, the calculation of phonon frequencies in a $2\times2\times1$ supercell of the $C2/c$ structure with 104 atoms and indicates that the structure is dynamically stable. This result precludes the possibility of a spontaneous polarization emerging in this phase through another dynamic instability. 

In passing, we note that while the structures we predict for Sr$_3$(VO$_4$)$_2$ and Ba$_3$(VO$_4$)$_2$ correspond to the lowest energy structures, for Ca$_3$(VO$_4$)$_2$, we find the experimentally reported $R3c$ structure with 7 f.u. to be lower in energy. However, given the closer packing of ions in the palmierite structure, it is possible that this structure may be stabilized in intermediate pressures between the observed $R3c$ and $C2/m$ structures \cite{Grzechnik2002}.
 

\subsection{Landau free energy and order parameter directions \label{sub:landau}}

As discussed in the previous section, the lowest energy commensurate structure hosts a combination of  $T_1^-$ and $T_3^-$ irreps, and has the $C2/c$ symmetry. In this section, we build the lattice free energy as a function of structural distortions that transform as $T_1^-$ and $T_3^-$ to explain the stabilization of this specific ground state, as opposed to others (for example $C2/m$) with a different direction of the same order parameters. 

We denote the two components of the $T_3^-$ order parameter as $(a,b)$. Note that since the parent group is trigonal, the two components are not symmetry equivalent, and $T_3^-(a,0)$ and $T_3^-(0,b)$ lead to different symmetries. 
The most general invariant polynomial up to 4$^{th}$ order in $(a,b)$ is
\begin{equation}
    \mathcal{F}(a, b) = \alpha(a^2+b^2) + \beta(a^2+b^2)^2 .
\end{equation} 
Since the free energy expression is depends only on  in $a^2+b^2$, it  does not have a preferred direction in the phase space spanned by $a$ and $b$, and hence cannot distinguish between $C2/c$ ($T_3^-(a,0)$), $C2/m$ ($T_3^-(0,b)$), or $P\bar{1}$ ($T_3^-(a,b)$) phases. It is either the higher order terms, or coupling with the other order parameter $T_1^-$ that breaks this symmetry. 
The free energy for $T_1^-(c)$ distortion itself is
\begin{equation}
    \mathcal{F}(c) = \alpha' c^2 + \beta{'}c^4
\end{equation}
which is trivial. However, when these two irreps coexist, the Landau free energy expression involves additional cross terms:
\begin{equation}
\begin{split}
    \mathcal{F}(a, b, c) = & \alpha(a^2+b^2) + \alpha{'}c^2 + \beta(a^2+b^2)^2 + \\
    & \beta{'}c^4 + \gamma(a^2+b^2)c^2 + \lambda (a^3-3ab^2)c
\end{split}
\end{equation}
If we write $(a,b)$ in terms of an amplitude ($r$) and an angle ($\phi$) where $(a, b) \equiv (r\cos{\phi}, r\sin{\phi})$, the free energy expression becomes
\begin{equation}
\begin{split}
    \mathcal{F}(r, \phi, c) = &\alpha r^2 + \alpha{'} c^2 + \beta r^4 + \beta{'} c^4 + \gamma r^2c^2 + \\
    & \lambda r^3c \cos(3\phi)
\end{split}
\end{equation}
which depends on the direction of $T_3^-$. This free energy has six minima with equal energy at $\phi=n\pi/3$ for integer $n$. Even though the sign of $c$ is opposite in three of these minima, all of them are different domains of the same $C2/c$ phase. Thus, the form of fourth-order free energy determines the lowest-energy state to be necessarily $C2/c$, and only higher-order terms or possibly strain coupling can change the ground-state symmetry. 


%

\subsection{\label{sub:bondvalence} Bond valence calculations}

The discussion so far focuses on different structures' stabilities using first principles calculations, but does not provide any insight on the driving force behind these transitions. Octahedral rotations in perovskites oxides are driven by the underbonding of $A$-site cations, and can be predicted with good accuracy using simple bond valence approaches \cite{Lufaso2004}. Given the highly ionic nature of the $M_3$(VO$_4$)$_2$ compounds we study, bond valence sums might be relevant in explaining the tetrahedral-tilting structural transitions in these compounds. Hence, we employ the bond valence approach to predict the structural instabilities in these compounds as well.

In the bond valence approach, empirical parameters for each cation-anion pair are used to estimate the valence associated with a bond, and the valences of all bonds that a cation makes are summed to obtain a total bond valence of the cation. The crystal structure that optimizes the bond valence of all cations (i.e. gives the closest value to their nominal valences) is the most favorable one for a compound.  
The bond valence $v_{ij}$ of a bond between atoms $i$ and $j$ that is of $R_{ij}$ in length is given by
\begin{equation}
       v_{ij} = \exp{\left[(R_0-R_{ij})/b\right]},
\end{equation}
where $R_0$ is an empirical value of the expected bond length, and $b$ = 0.37 {\AA} is an empirical constant \cite{Brese1991}.
We obtain the valence $V_i$ of an atom $i$ by summing the individual bond valences $v_{ij}$ of all bonds involving that atom:
\begin{equation}
    V_i = \sum_j v_{ij}.
\end{equation}

The discrepancy factor $d_i$ is defined as the difference of the calculated bond valence $V_i$ and the nominal valence of a cation (5 for V$^{5+}$), in other words, $d_i = V_i - V_{\text{formal}}$.
The sign of the discrepancy factor determines whether an ion is overbonded or underbonded. The root mean square of all the discrepancy factor in a structure is called the Global Instability Index (GII) \cite{Brown2009}, which can be used to estimate the overall stability of the structure:
\begin{equation}
    \text{GII} = \left(\sum_i^N \frac{d_i^2}{N}\right)^{1/2} .
\end{equation}

\begin{table}[]
    \centering
    \begin{tabular}{c c c c}
    \hline
    \hline
    $R\bar{3}m$ &  Ca$_3$(VO$_4$)$_2$ & Sr$_3$(VO$_4$)$_2$ & Ba$_3$(VO$_4$)$_2$ \\
    \hline
    M1 & 2.09 & 2.18 & 2.24\\
    M2 & 1.85 & 1.94 & 2.07\\
    V & 5.16 & 5.13 & 5.08\\
    O1 & 1.99 & 1.95 & 1.92\\
    O2 & 2.06 & 2.11 & 2.15\\
    \hline
    GII (vu) & 0.09 & 0.12 & 0.15 \\
    \hline
    \end{tabular}
    \caption{Bond valence and global instability index (GII) of $M_3$(VO$_4$)$_2$ in high-symmetry $R\bar{3}m$ structure.}
    \label{tab:bv1}
\end{table}

\begin{table}[]
    \centering
    \begin{tabular}{c   c   c}
    \hline
    \hline
    $C2/c$ &  Ca$_3$(VO$_4$)$_2$ & Sr$_3$(VO$_4$)$_2$ \\
    \hline
    M1 & 2.19 & 2.21 \\
    M2 & 1.92 & 2.01 \\
    V & 5.07 & 5.09 \\
    O1 & 2.06 & 2.00 \\
    O2 & 2.01 & 2.09 \\
    O3 & 2.17 & 2.04 \\
    O4 & 1.98 & 2.18 \\
    \hline
    GII (vu) & 0.11 & 0.12 \\
    \hline
    \end{tabular}
    \caption{Bond valence and global instability index of $M_3$(VO$_4$)$_2$ in $C2/c$ structure.}
    \label{tab:bv2}
\end{table}

The bond valence parameter and the GII have been proven to be very effective tools to explain structural stability of crystals where structural transition leads to a change of local bonding environment \cite{Brown2009,Brown1978,Brese1991,Woodward1997}. 
The bond valence and the GII of the $R\bar{3}m$ and $C2/c$ orthovanadate structures are listed in Tables \ref{tab:bv1} and \ref{tab:bv2}.
Although the GII of high-symmetry and low-symmetry structures are very comparable, interesting patterns emerges while exploring their individual bond valences.

The vanadium ions are overbonded in all the structures, and as the size of the $M$ atom decreases, the bond valence increases for the high-symmetry structure.
The $M$ atom at the Wyckoff position $a$ is consistently underbonded, and the bond valence decreases with decreasing atom size.
These two atoms becomes more stable in the low-symmetry structure.
However, the $M$ atom at the Wyckoff position $c$ is overbonded in the high-symmetry structure, and becomes further overbonded in the low-symmetry structure.
This suggests that the bond valence of the V atom and the $M$ atom at cite $a$ play a bigger role in the stabilization of the orthovanadate structure. 
Overall, the bond valences of the cations do not predict the transition from $R\bar{3}m$ to $C2/c$. The difference of the bond valences of the two phases of Sr$_3$(VO$_4$)$_2$ is essentially zero, and that of Ca$_3$(VO$_4$)$_2$ just 0.02 with the wrong sign: the bond valence is larger in the lower-symmetry structure, which means that the structural distortions would not exist, which is against the DFT results presented earlier. 

One of the effects that the bond valence model does not capture is the anion-anion repulsion. To incorporate this, we calculated the anion-anion repulsion energy by summing up the reciprocal of all the oxygen-oxygen distances in different structural phases. This term is analogous to the Coulomb repulsion between the O ions treated as point charges. Although this sum diverges, the difference of its value in two different structures, truncated at a large distance, can in principle be used for comparison between structures. 
%

The results shown in detail in the Supplemental Material \cite{Note1}  show that the structure with $C2/c$ symmetry higher in repulsion energy for Ca$_3$(VO$_4$)$_2$, and  lower for Sr$_3$(VO$_4$)$_2$, which is inconsistent with an instability emerging in both of them.
Overall, this suggests that the stabilization of the $C2/c$ structure is not caused by anionic repulsion, and maybe the bond valence of the V atom and the $M$ atom at cite $a$ play a very important role in predicting the structures.

These observations pose an open question about the driving force of the structural distortions in $M_3$(VO$_4$)$_2$. The large band gap of these compounds (discussed in the next section) implies that complicated covalency effects are likely negligible, and therefore the structural instabilities should be well explained by the bond valence approach, which is clearly not the case. One possibility that explains this discrepancy may be the bond valence parameters used, which can always be improved by the inclusion of more compounds with V$^{5+}$ into the data sets used to fit those parameters.

\subsection{Electronic structure \label{sub:electronic}}

\begin{figure}
    \centering
    \includegraphics[width=0.98\linewidth]{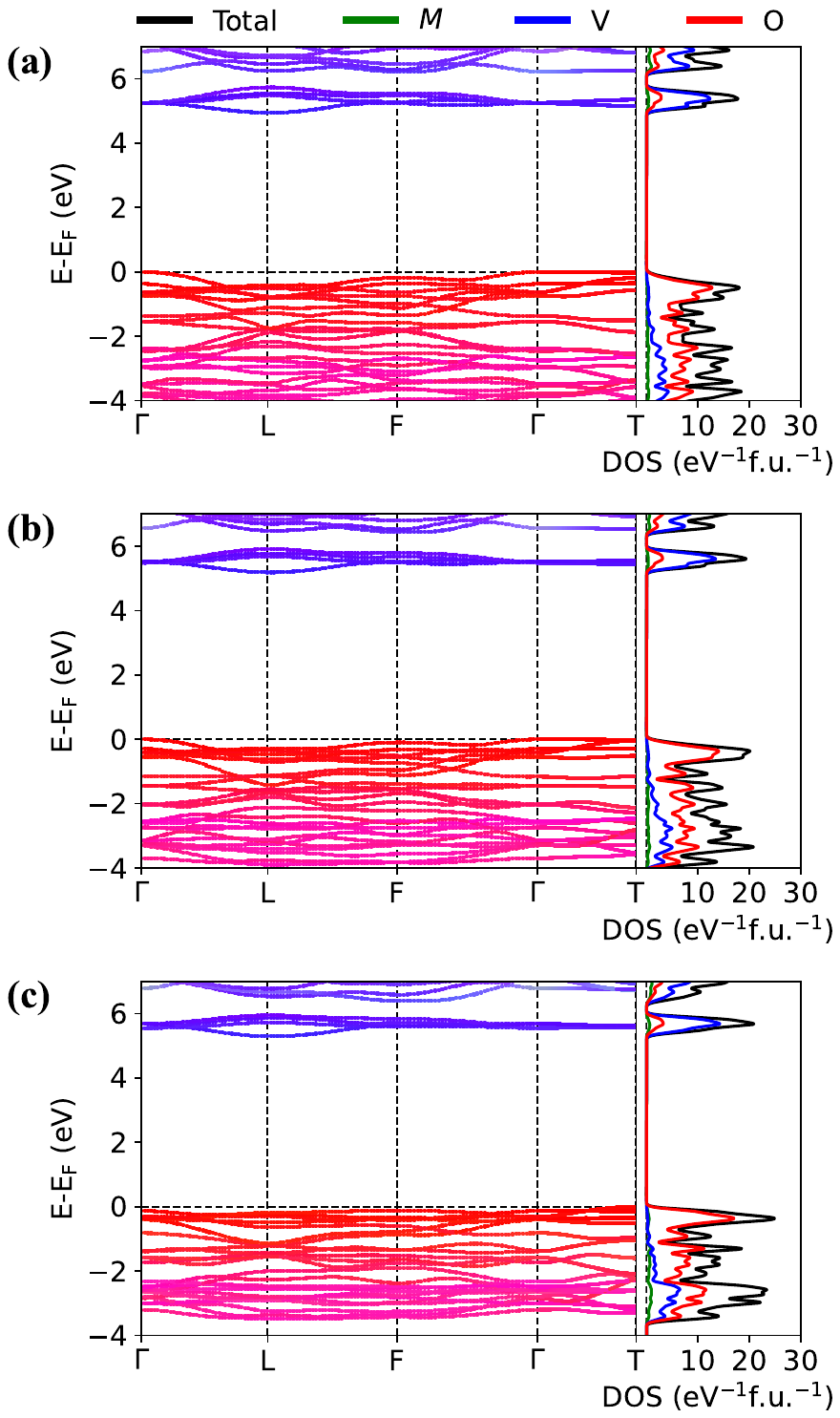}
    \caption{Band structure and density of states for $R\bar{3}m$ $M_3$(VO$_4$)$_2$ obtained using HSEsol exchange correlation function. The atomic characters of the electronic wavefunctions were obtained from atomic spheres, renormalized to account for interstitial densities, and denoted with  green ($M$), blue (V), and red (O) colors. Panels a, b, and c correspond to Ba-, Sr-, and Ca- compounds respectively.}
    \label{fig:band}
\end{figure}

\begin{figure}
    \centering
    \includegraphics[width=0.98\linewidth]{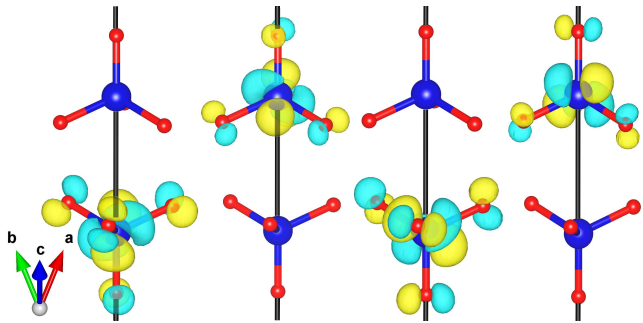}
    \caption{Maximally localized Wannier orbitals of the four lowest conduction bands of Sr$_3$(VO$_4$)$_2$. The shape of these orbitals suggests that they are a hybrid of the vanadium $d$ and oxygen $p$ orbitals, which can also be seen in Fig.~\ref{fig:band} density of states.}
    \label{fig:wannier}
\end{figure}

Despite the large number of studies on the electronic structure of V$^{5+}$ containing transition metal oxides, to the best of our knowledge, there is no first principles study of the electronic structure of $M_3$(VO$_4$)$_2$. In order to clarify the electronic structure of these compounds, and the types of orbitals that contribute to the bands near the Fermi level and hence determine the optical properties, we perform band structure and Wannier function calculations of all three $M_3$(VO$_4$)$_2$ compounds. 

For simplicity, we explore the electronic properties in the parent $R\bar{3}m$ phase and keep the notation of the $k$-points in the Brillouin zone consistent among the compounds. 
In Fig.~\ref{fig:band}, we present the atom-projected electronic band structure and density-of-states (DOS). The DOS reveals that the valance band is dominated by oxygen, while the conduction band is largely vanadium. All three compounds display sizable bandgaps larger than 4.5~eV even with the underestimation of DFT. The bandgap increases with increasing $M$-ion radius, and is largest in Ba$_3$(VO$_4$)$_2$. The separation of the oxygen $p$ and vanadium $d$ orbitals is larger in these orthovanadates than that observed in AVO$_3$ perovskites, and even larger than that in SrNbO$_3$ with the more electropositive Nb ion \cite{Paul2019Vanadate, Paul2020, Park2020}. This difference, as a result, is likely a result of the higher valence of V in this compound, rather than an effect of the crystal structure and the different coordination environment. 

The bottom of the conduction band consists of parabolic minima at the $L$ points, which display some dispersion despite the disconnected arrangement of the VO$_4$ tetrahedra. These minima are part of 4 isolated V-induced bands. The Wannier orbitals built from this isolated manifold, shown in Fig.~\ref{fig:wannier}, display $e_g$ character (similar to d$_{3z^2-r^2}$ and d$_{x^2-y^2}$ orbitals in perovskites) with lobes extended towards oxygen anions. This is consistent with the tetrahedral environment of the V ions, which causes an opposite $t_{2g}$--$e_g$ splitting to that in octahedral coordination environment. There is a non-negligible hybridization between the V orbitals and the O $p$ orbitals, as seen in both the Wannier functions, and the DOS.

\section{\label{sec:conclusion} Conclusions}

By performing first principles evolutionary structure prediction and lattice dynamics calculations, supported by a group theoretical analysis, we elucidated the crystal structures and possible low temperature or metastable phases of $M_3$(VO$_4$)$_2$ compounds. We showed that the higher valence of vanadium leads to a tetrahedral coordination of oxygen ions around the vanadium centers. These oxygen tetrahedra rotate to lower the energy of the crystal structure, which is similar to the ubiquitous octahedral rotations in perovskites. The lowest energy state with $C2/c$ symmetry of Sr and Ca compounds is stabilized through the coupling of a strong unstable $T_3^-$ phonon with another $T_1^-$ mode. 
We also find that even though there is a polar $\Gamma$ point instability, it is too weak to give rise to a polar phase at room temperature, and is suppressed by the other instabilities. This suggests that the reported dielectric anomaly in Sr$_3$(VO$_4$)$_2$ likely has an extrinsic origin. 

All three compounds we studied have large band gaps, which make them transparent throughout the visible range, and opens the way for optics-related applications like other orthovanadates \cite{Nakajima2015}. 

In summary, our study shows that the palmierite-derived orthovanadates have a rich structural energy landscape with multiple instabilities and competing phases and, so, this structural family is a prime target for designing new materials with novel functionalities. 

\acknowledgements

This work is supported by the Office of Naval Research Grant No. N00014-20-1-2361. We acknowledge useful discussions with R. Engel-Herbert.

\end{document}